# Placental contractions in uncomplicated pregnancies

**Authors:**

Louise Dewick[1*], Amy L Turnbull[2], Kate F Walker [1], Nia W Jones [1], George Hutchinson [2], Christopher Bradley [2], Taqwa Ferdous [3], Aisha Razzaque [3], Ruizhe Li [4], Xin Chen [4], Grazziela Figueredo [4], Craig Platt [5], Cesar Peres [5], Penny Gowland [2]

**Affiliations**

[1] Centre for Perinatal Research, University of Nottingham, Nottingham, UK

[2] Sir Peter Mansfield Imaging Centre, University of Nottingham, Nottingham, UK

[3] Department of Obstetrics and Gynaecology, Nottingham University Hospitals NHS Trust, Nottingham, UK

[4] School of Computer Science, University of Nottingham, Nottingham, UK

[5] Department of Histopathology, Nottingham University Hospitals NHS Trust, Nottingham, UK

*Corresponding author

Email: louise.dewick@nottingham.ac.uk (LD)

LD and ALT are Joint Senior Authors

# Abstract

In 2020 we first described placental contractions, and we have now undertaken a study to characterise them and seek features that might automatically separate them from uterine contractions. We recruited 36 healthy pregnant women to undergo magnetic resonance imaging (MRI) between 29 and 42 weeks of pregnancy in a single-centre, prospective, observational study. Participants had fetal ultrasound to confirm normal growth. Dynamic MRI was acquired for between 15 and 32 minutes using respiratory triggered, multi-slice, single shot, gradient echo, echo planar imaging covering the whole uterus. All participants had a live birth of a healthy baby weighing over the $10^{th}$ centile for gestational age and none developed any associated conditions of placental dysfunction e.g. pre-eclampsia, or severe maternal or fetal villous malperfusion on placental histopathology.

Any visible contractions were recorded for all participants who completed their MRI scan and placental contractions occurred in at least 60% of our healthy pregnant population with a median frequency of approximately 2 per hour, and a median duration of 2.4 minutes.

Contractions involving a decrease in placental volume of >10% were classified as either placental or uterine by visual observation. Placental contractions occurred more frequently than uterine contractions ($p=0.0061$), were associated with a larger increase in the surface area of the uterine wall *not* covered by the placenta ($p=0.0015$), placental sphericity ($p<0.0001$) and longer duration ($p=0.0151$). All contractions led to an increase in the MRI parameter $R_2^*$ in the placenta. There was large variation both between participants and between contractions from the same individual, in terms of time course and contractions features, with no apparent change across the gestational age range studied, although the largest fractional volume changes were detected at early gestation.

We found that the change in sphericity could provide a marker for separating placental from uterine contractions causing >10% volume decrease.

# Introduction

A healthy placenta (Figure 1) is key to a successful pregnancy outcome, but much is unknown about the function of this vital organ. Placental dysfunction can lead to morbidity and mortality for both the mother and baby, including antepartum stillbirth, which affects over 1 in 250 pregnant women in the UK [1]. Estimates suggest approximately 35% of these stillbirths are attributable to placental dysfunction [2]. Methods of detecting developing placental dysfunction are largely indirect, relating to the growth and activity levels of the fetus. However, ultrasound assessment of fetal growth may be inaccurate or falsely reassuring and maternal perception of fetal movements is subjective. This has led to a focus on finding new reliable methods to detect placental dysfunction.

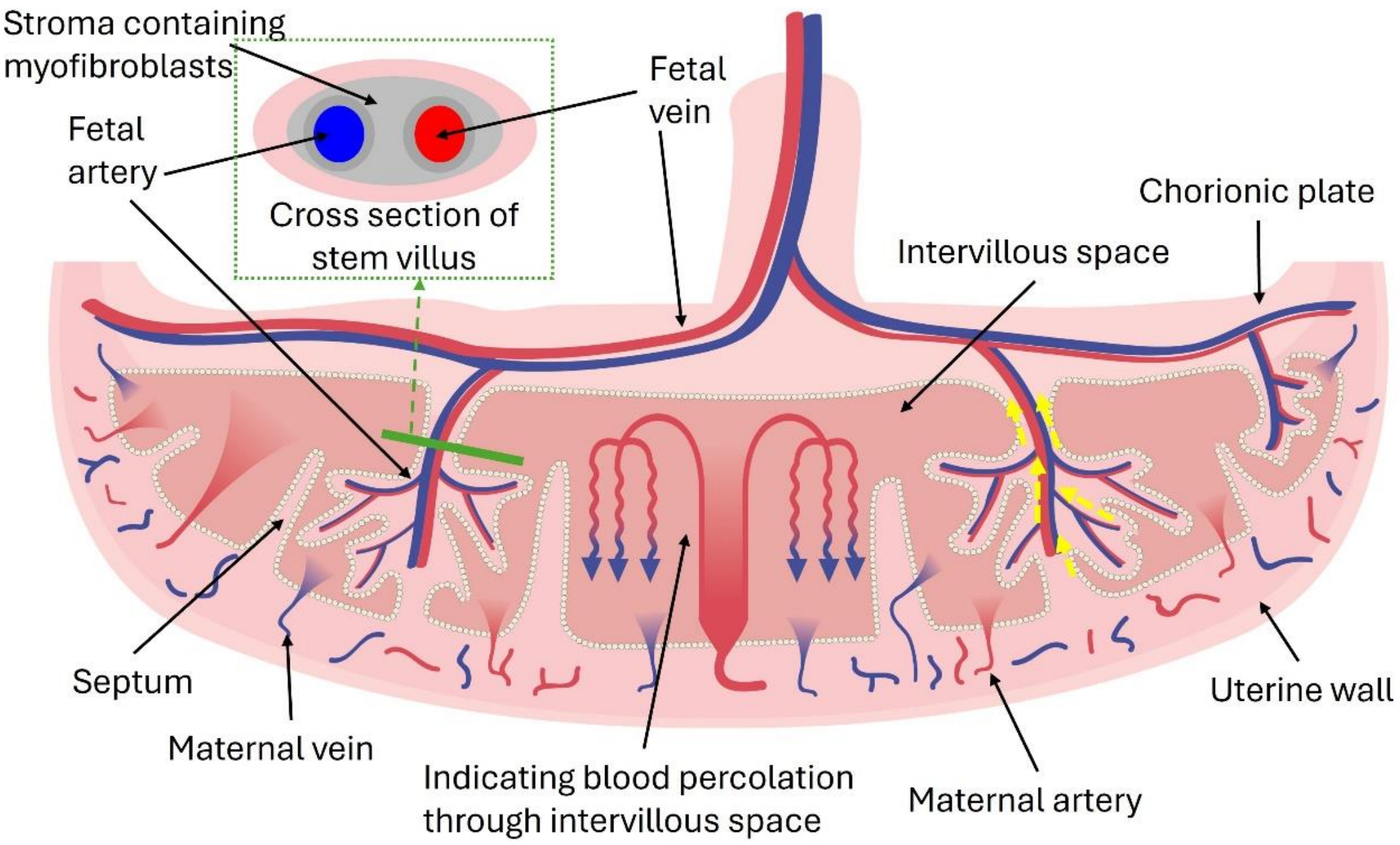

**Fig 1. A schematic illustrating placental structure.** Both arteries and veins contain smooth muscle in their vessel walls, but large stem villi (over 300µm), also contain myofibroblasts (smooth muscle cells) in the stroma surrounding the vessels, (termed Perivascular Contractile Sheath or Extravascular Contractile System). Stromal myofibroblasts are arranged parallel to the longitudinal axis of the stem villi and thus have the potential to induce longitudinal contraction (Demir 1997) (depicted by yellow dashed arrows), which could force blood out of the intervillous spaces via the maternal veins.

Recently, advances in magnetic resonance imaging (MRI) have allowed researchers to explore markers of placental function. In particular, placental $R_2$*, an MRI parameter that depends on multiple aspects of tissue structure including movement, but particularly concentration and distribution of deoxygenated blood, is known to increase in conditions associated with placental dysfunction [3]. $R_2$* is frequently referred to by its reciprocal $T_2$* and is also known to increase with gestational age [4-7].

Uterine contractions have long been observed in dynamic MRI scans of the uterus in pregnancy [8], and have generally been assumed to be Braxton Hick's contractions, involving the entire uterine wall. Sinding et al [3] noted the occurrence of spontaneous non-labour contractions (a change in shape and cross sectional area of the uterus) on 5 minute dynamic (cine) $R_2$* MRI scans in 8 out of 56 cases, with the change in shaped being immediately followed by a significant reduction in the T2* MRI signal which they associated with a reduction in placental oxygenation. In 2020 we reviewed contractions observed on dynamic $R_2$* weighted MRI and observed a new physiological phenomenon: the placental contraction or 'utero-placental pump', where the placental bed contracts independently of the rest of the uterine wall [8]. These placental contractions were defined as contractions largely isolated to the placenta and underlying uterine wall. In 2025 a large study reviewed over 800 MRI dynamic data sets of around 5 minutes duration for the presence of transient reductions in T2*, which were interpreted as contractile activity. They concluded episodes of uterine contractility occurred in 19% of the cases analysed, which included both healthy and compromised pregnancies.

However, they only assessed T2* changes and did not consider any accompanying changes in the shape or morphology of the uterine wall or placenta [9].

Despite this growing body of work, the basic characteristics of the different types of contractions need to be determined. Furthermore, currently no system exists to classify whether contractile events seen on MRI during pregnancy are uterine or placental in origin. If the relevance of contractions in healthy development and compromised pregnancy is to be investigated, then a systematic and ideally objective approach is required to separate them.

The aim of this study was to use MRI to determine the magnitude, duration and frequency of placental or uterine contractions in healthy pregnancies, classifying the contractions visually as uterine or placental. We also measured basic morphological and MRI features with the goal of identifying signatures that will allow automatic discrimination of placental from uterine contractions in future. This is a part of a larger multidisciplinary study with the goal of finding markers of placental stillbirth risk.

# Methods

### Ethics statement

Participants receiving antenatal care at Nottingham University Hospitals NHS Trust were recruited for the SWIRL (Stillbirth, When Is Risk Low?) study, and written, informed consent was obtained from each participant. The study was approved by the East Midlands–Leicester Central Research Ethics Committee (23/EM/0052).

# Participant characteristics

Between 10$^{th}$ July 2023 and 29$^{th}$ February 2024, we recruited 36 women experiencing healthy, uncomplicated pregnancies to attend a single MRI scan between 29$^{+0}$ and 42$^{+6}$ weeks of pregnancy. As it was a pilot study, no formal sample size calculation was performed.

Inclusion criteria were singleton pregnancy, aged 18-50 years old, ability to give informed consent and the absence of risk factors for fetal growth restriction according to the Saving Babies Lives Care Bundle version 2[10]. Exclusion criteria included women with a body mass index (BMI) of 45 kg/m$^2$ or over, claustrophobia, any contraindication to MRI scans, a baby with a lethal congenital abnormality and women without an ultrasonographic dating scan before 22 weeks' gestation.

All participants underwent an ultrasound assessment of fetal biometry, amniotic fluid volume, umbilical artery Doppler and uterine artery Doppler. To be considered low-risk they needed to have an estimated fetal weight above the 10$^{th}$ centile for gestational age using Intergrowth 21 charts[11], an amniotic fluid index greater than 5cm and umbilical/uterine Dopplers under the 95$^{th}$ centile for gestation [12, 13]. Clinical outcome data was collected retrospectively using the electronic patient record.

# MRI

Participants were positioned at a lateral tilt, and scanned using a 3T Philips Ingenia MRI scanner, with one dStream anterior body coil placed anterior to the participant and the posterior bed coil. If localiser scans revealed a posterior placenta with low signal to noise ratio across it, the participant was repositioned with the anterior body coil placed posteriorly (N=2 cases). Initially, multimodal MRI was performed to assess placental blood flow (data not reported here).

Dynamic MRI (a time course of multislice volume datasets) was then acquired for between 15 and 32 minutes, with the length of data acquisition depending on the mother's comfort, using Echo Planar Imaging (EPI). EPI is a very rapid MRI method that allows dynamic imaging of morphological changes in the placenta, which is $R_2^*$ weighted.

Multislice volume datasets were acquired trans-axially to the mother, with the field of view covering the entire uterus. The sequence used was multi-slice, single shot, gradient echo EPI (echo time 25ms, voxel size 2.4x2.4x6 $mm^3$, 32 slices, slice gap 4 mm, field of view 400x400x316 mm, Sense 3). The acquisition was respiratory triggered to expiration (trigger delay 200ms, minimum repetition time 9s, mean repetition time 15s).

A relatively large interslice spacing (10 mm) was used to achieve full coverage across the whole uterus with relatively few slices; this interslice spacing has minimal effect on the accuracy of the volume estimates for these smooth objects. Full volume coverage was required to ensure that we accurately measured changes in placental and uterine volumes/areas, regardless of through plane motion between time points. Reducing the number of slices allowed us to limit the acquisition to the quiescent phase of the respiratory cycle, to ensure minimal motion through and between each volume, which simplified segmentation of the uterus and placenta. Fewer slices also reduced the acoustic burden on the mother and fetus. A slice gap (rather than increased slice thickness) was used to achieve this interslice spacing, to reduce slice cross talk due to movement and improve spatial resolution by reducing partial volume effects. The slices were acquired sequentially so that any residual motion between slices would be easier to interpret.

## Data Analysis

Images were automatically segmented to estimate placental volume, fetal and fluid volume, placental bed area and remaining uterine wall (non-placental bed) area as shown in Figure 2. Segmentation was performed by nnU-Net[14] a convolutional neural network trained on 169 manually segmented volumes from 39 individuals. To evaluate segmentation performance, a 5-fold cross-validation at the subject level was conducted. The average Dice Coefficient across all volumes was 0.8134 ± 0.0560 for placenta and 0.9542 ± 0.0158 for the non-placental uterus. These results indicate good segmentation performance for both structures, with particularly high accuracy for the non-placental uterus.

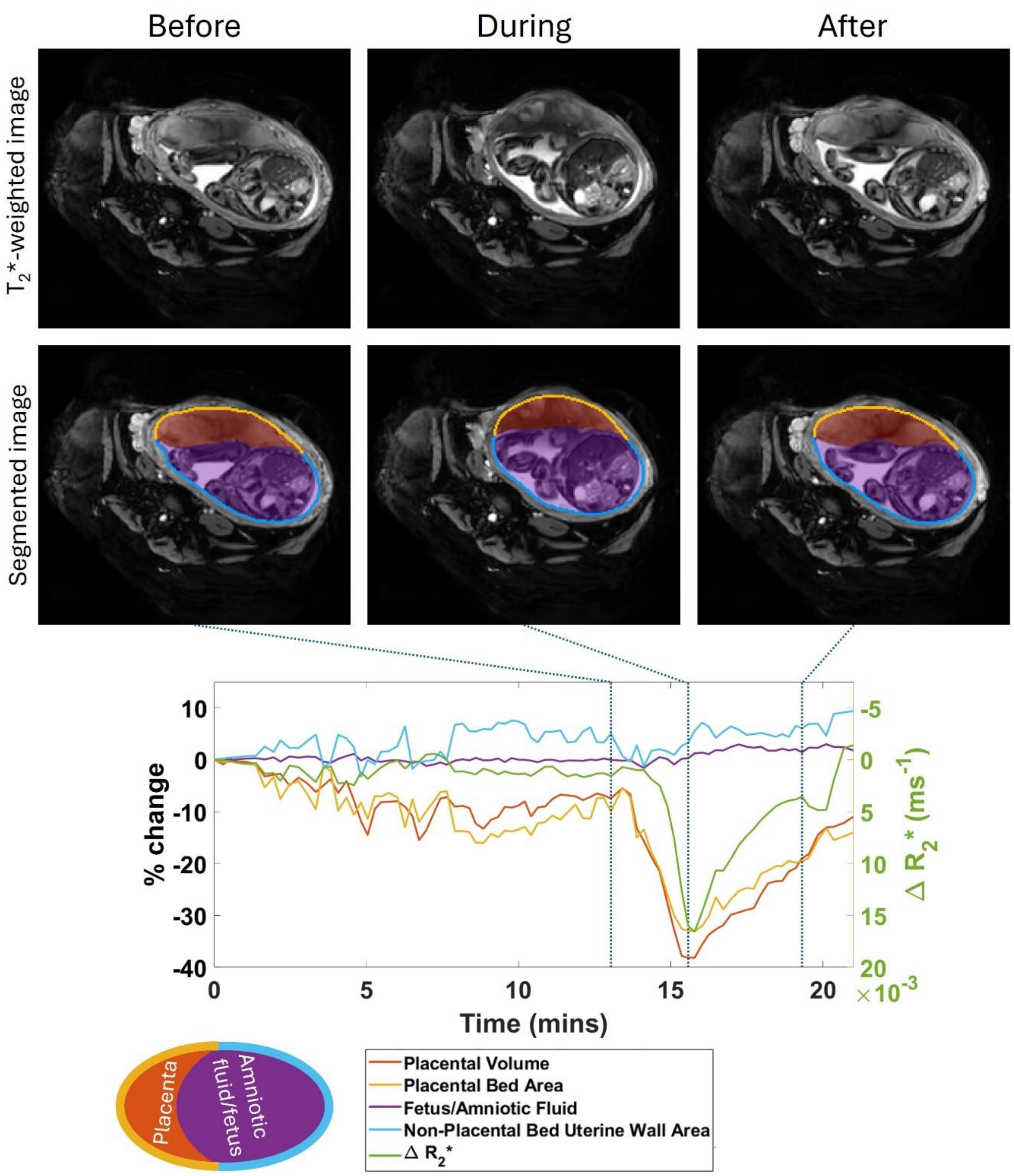

**Fig 2: Uterine and placental changes during an example placental contraction.** The top row shows axial MRI images at selected times points before, during and after a placental contraction (indicated by vertical lines on the graph below) with segmentations shown below. Changes in placental and non-placental volumes, wall areas and placental $R_2$* (all measured across the whole volume of the uterus, not just the single slice shown) are plotted underneath. The legend indicates the colours used for the lines in the plots and the regions indicated in the segmentation.

The segmentations were used to create time courses of measured volumes ($V(t)$) and areas ($A(t)$), which were smoothed using a moving average filter with a span of 5 timepoints. Placental sphericity at each time point was calculated using

$$Sphericity(t) = \frac{\pi^{\frac{1}{3}}(6V(t))^{\frac{2}{3}}}{A(t)}.$$

(yielding a dimensionless value between 0 and 1). The change in $R_2$* across the placenta mask at each time point was calculated using

$$\Delta R_2^*(t) = -\frac{1}{TE} ln\left(\frac{S(t)}{S_b}\right)$$

where $S(t)$ is the average signal intensity in the placenta at each dynamic, $S_b$ is the initial baseline value at the earliest rest volume (which was not the initial volume if the data acquisition started during a contraction) and TE is the echo time.

Time courses were plotted as a percentage of the earliest rest volume. The maximum percentage changes in volumes and areas as well as the change in $R_2$* during the contraction were calculated relative to a local baseline. This local baseline was approximated as being linear between the value at the last rest volume before the contraction to the value at the first rest volume following it (vertical lines 1 and 3 on Figure 2). In cases where the start/end of a contraction occurred outside the scan duration, the local baseline was assumed to be constant at the value at the end/start. The duration of contraction was reported as the time between half maximum change in volume.

Contractions were visually identified by inspecting both the cine series of dynamic images across multiple slices and plots of the changes in volume and area measurements together. After training and familiarization with the data (working with TF, AR and ND), two raters (PAG: 30 years' experience of placental MRI and ALT: 3 years' experience of placental MRI) identified contractions in all time courses, considering both the cine image time series (over multiple slices) and the volume/area time courses (Figure 3, and S1a and b). Both observers agreed on the occurrence of all contractions though they did not always agree on observed start and end times, particularly since the end could show an extended tail as the contraction relaxed (Figure 2). The number of contractions detected was divided by the time over which data was acquired, to estimate the mean number detected per hour.

Given previous knowledge that women experience contractions or 'tightenings' during pregnancy, sometimes referred to as 'Braxton Hicks' and assumed to be uterine contractions, two observers (PAG and ALT individually and then in consensus), classified contractions into two types based on dynamic morphological changes observed in the cine image time series and volume time courses (Figure 3).

- Placental contractions were defined as those generating an apparent shortening of the placental bed in the 2D images, an observed thinning of much of the non-placental bed section of the wall, and a change in placental shape particularly at the periphery of the placenta. Generally, the periphery of the placenta became more square and less tapered in profile and, if the marginal sinus was apparently visible, it usually disappeared.
- Uterine contractions showed little change in placental shape or size on the images, beyond sometimes becoming apparently thinner, with thickening of the uterine wall which could be either (a) local but not under the placenta (red arrows in Figure 3) or (b) uniform.

If both types of contraction were to occur simultaneously it is likely that they would be categorised as uterine by this process. We did this for all contractions but then separated out those with placental volume change >10%, as consensus was sometimes difficult to achieve for smaller contractions.

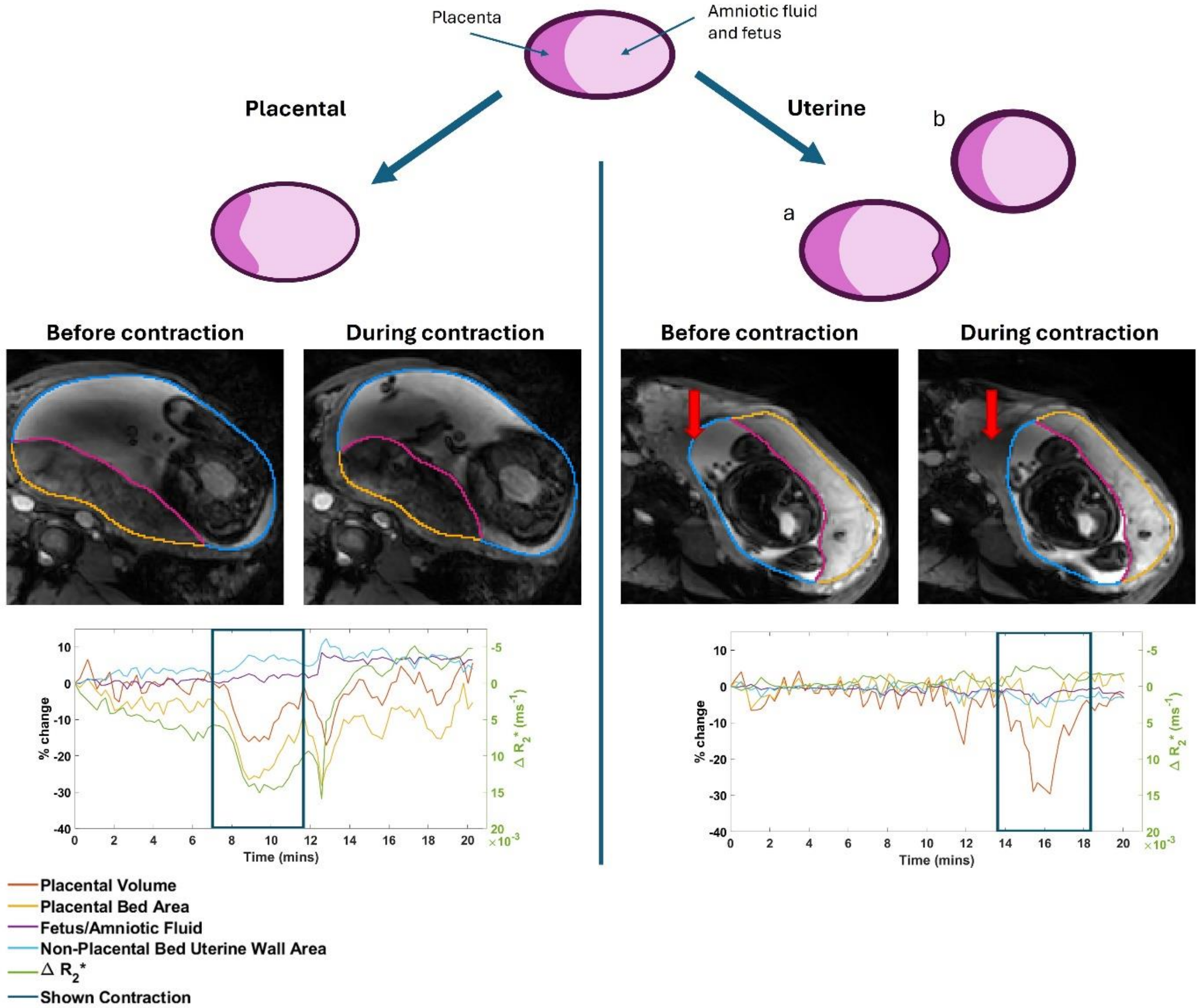

**Fig 3: The morphological changes used to visually categorise placental and uterine contractions.** Placental contractions are generally associated with an apparent shortening of the placental bed in the 2D images, a thinning of the wall not covered with the placenta, and a change in placental shape particularly at the periphery of the placenta. Uterine contractions are generally associated with little change in placental shape except some thinning, with thickening of the uterine wall that can be either (a) local (red arrows) or (b) uniform. The legend colours match Figure 2 but regions are not shaded to improve visualization, so the pink line indicates the amniotic surface of the placenta. The plots show the changes in volumes, areas and signal with the example contraction highlighted. Both time courses show additional contractions with <10% placental volume change: LHS from 14.7-16.6 mins and 17.8 – 20.3 mins, RHS from 10.8 – 13.2 mins.

Maternal motion events (which were rare) were detected in the time courses by thresholding the $R_2^*$ weighted images to create a mask of high signal maternal features, and then removing the uterine mask produced from the automatic segmentation. The difference between the current and previous mask was then summed over the whole image at each timepoint to give a maternal movement time course. A similar process was used to detect motion events of the fetus (which were more common). The high signal amniotic fluid was selected within the non-placental uterine volume mask using a simple threshold to produce a fetal movement time course. If the average maternal or fetal movement across all non-contracting timepoints was greater than 2% then it was considered a high movement scan.

## Statistical analysis

Differences between groups were compared using the Mann Whitney U test in GraphPad. Bonferroni correction was applied when comparing metrics between placental and uterine contractions.

# Results

## Participant Demographics

The demographics of our cohort are shown in Table 1. All 36 women recruited were healthy with no chronic medical conditions known to impact fetal growth. One participant withdrew a few minutes into the scan due to feeling unwell and was removed from further analysis. Participants were all non-smokers, aged under 40 years old and had a first trimester BMI below 35kg/m$^2$. None of the participants developed any conditions associated with placental dysfunction e.g. fetal growth restriction, pregnancy induced hypertension or pre-eclampsia.

**Table 1. Participant demographics**

| **Variable** | **Category** | **n (% of whole cohort)** |
| --- | --- | --- |
| Maternal Age | 18-24 years old | 1 (3%) |
| | 25-29 years old | 4 (11%) |
| | 30-34 years old | 18 (51%) |
| | 35-39 years old | 12 (34%) |
| Maternal first trimester BMI (Body Mass Index) | <20 kg/m$^2$ | 1 (3%) |
| | 20-24kg/m$^2$ | 15 (43%) |
| | 25-29 kg/m$^2$ | 14 (40%) |
| | 30-34 kg/m$^2$ | 5 (14%) |
| Primiparous | | 12 (34%) |
| Multiparous | | 23 (66%) |
| Previous miscarriages | | 11 (31%) |
| No previous miscarriages | | 24 (69%) |
| Ethnicity (self-reported) | White British | 29 (82%) |
| | White European | 3 (9%) |
| | Asian – Indian/Bangladeshi | 2 (6%) |
| | Black African | 1 (3%) |

Ultrasound assessment of fetal wellbeing showed that all fetuses in our cohort had umbilical artery and mean uterine artery pulsatility index values below the 95$^{th}$ centile for gestation at the time of their MRI scan. Their fetal growth and liquor volume measurements are shown in Table 2.

**Table 2**. **Ultrasound assessment results to demonstrate fetal wellbeing**

| Variable | Category | n (% of whole cohort) |
|---|---|---|
| Estimated fetal weight centile | 10$^{th}$-25$^{th}$ | 2 (6%) |
| | 26$^{th}$-50$^{th}$ | 16 (46%) |
| | 51$^{st}$-75$^{th}$ | 12 (34%) |
| | 76$^{th}$- 99$^{th}$ | 5 (14%) |
| Amniotic Fluid Index | 5 - 9.9cm | 4 (11%) |
| | 10 -14.9cm | 20 (57%) |
| | 15 -19.9cm | 4 (11%) |
| | 20 -25cm | 3 (9%) |
| | unknown | 4 (11%) |

## Birth outcomes

The mean gestation at birth was 279 days or 39 weeks and 6 days. The earliest birth was at 37 weeks and 1 day, and the latest birth was at 41 weeks and 5 days. There were no babies born preterm (< 37 weeks' gestation). Table 3 outlines the labour and birth outcomes: all babies in our cohort had normal Apgar scores, no evidence of intrapartum hypoxia and no evidence of significant placental dysfunction on histopathological examination.

**Table 3. Labour and birth outcomes**

| Variable | Category | n (% of whole cohort) |
|---|---|---|
| Onset of labour | Spontaneous | 19 (54%) |
| | Induced | 9 (26%) |
| | No labour (elective caesarean) | 7 (20%) |
| Mode of birth | Spontaneous vertex | 16 (46%) |
| | Assisted/Instrumental | 3 (9%) |
| | Elective Caesarean | 7 (20%) |
| | Emergency Caesarean | 9 (25%) |
| Fetal Sex | Male | 15 (43%) |
| | Female | 20 (57%) |
| Birthweight centile | <3rd centile | 0 (0%) |
| | <10th centile | 0 (0%) |
| | ≥10th and ≤50th | 8 (23%) |
| | >50$^{th}$ and <90th | 19 (54%) |
| | >90th centile | 8 (23%) |
| 5-minute Apgar score | <3 | 0 (0%) |
| | <7 | 0 (0%) |
| | >7 | 35 (100%) |

| Placental histopathology | Severe features of Maternal Arterial Malperfusion | 0 (0%) |
|---|---|---|
| | Severe features of Fetal Villous Malperfusion | 0 (0%) |

Of the nine women who underwent induction of labour, five were induced due to reduced fetal movements at term, two had post-dates pregnancies, one had an unstable lie which became cephalic, and one participant had kidney stones requiring a nephrostomy. In the three women who underwent assisted vaginal/instrumental birth, the indication for one participant was delay in the second stage of labour, and for the other two there were fetal heart rate concerns. In the nine women who underwent emergency caesarean birth, one requested caesarean during labour, two had an unsuccessful induction of labour, three had delay in the first stage of labour, two had pathological fetal heart rate patterns and one had umbilical cord prolapse.

## MRI

Contractions of the uterus/placenta were observed in all participants who completed scanning occurring sporadically at a median (lower quartile, upper quartile) rate of 6.3 (3.9,9.0) contractions per hour, with 8.4 (5.1,14)% placental volume decrease, 1.6 (1.1,2.2) minutes' duration and $R_2^*$ increase of 0.00089 (-0.00096,0.0033) $ms^{-1}$; values reported are (median (lower quartile ,upper quartile) in all cases.

Figure 4a shows that there was considerable variation; some subjects showed one deep, long contraction (35% volume decrease in Figure 2), whilst others showed bursts of smaller contractions (Figure 3).

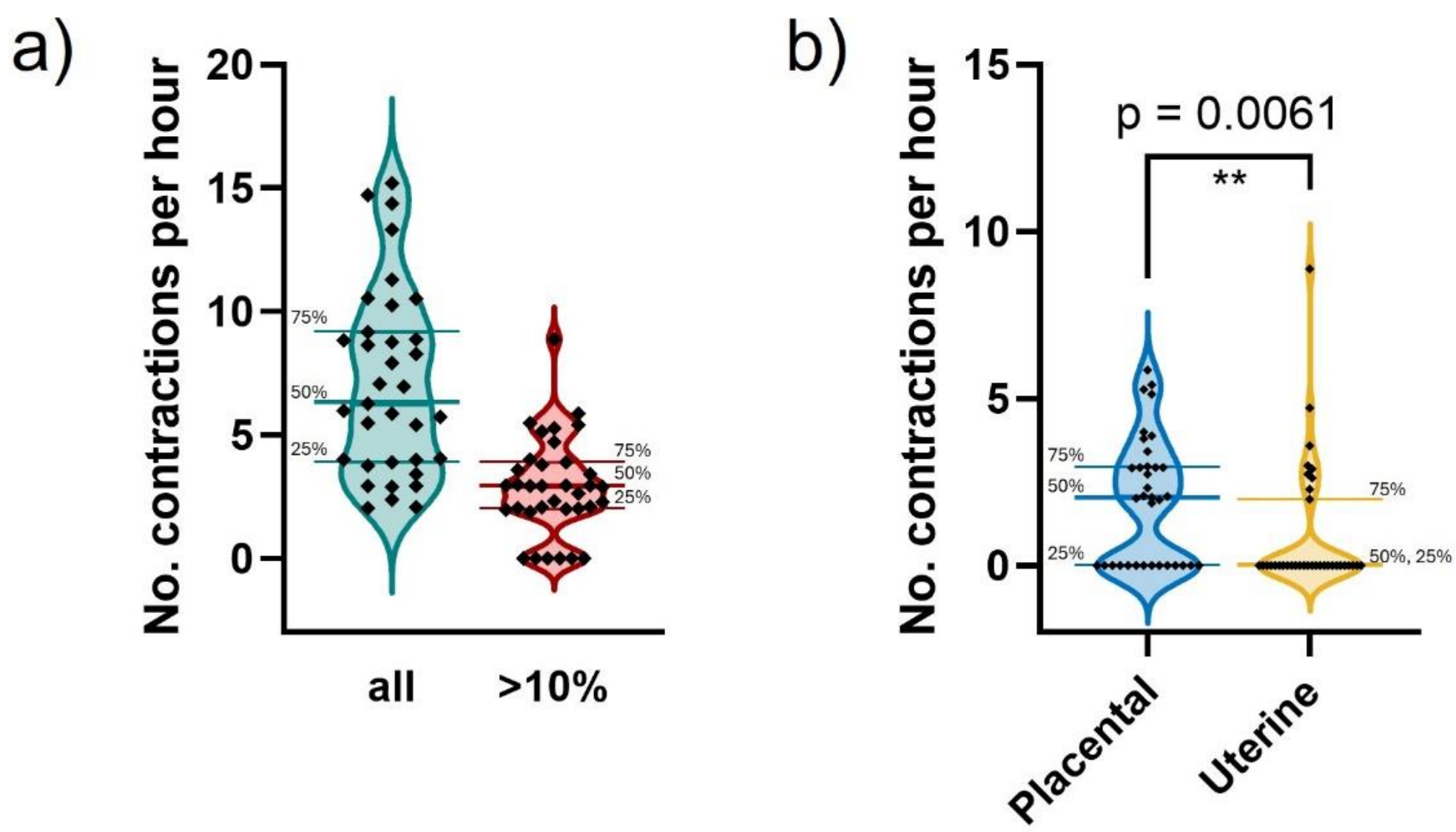

**Fig 4: Number of contractions detected per hour** (calculated from number detected divided by observation time). Median and interquartile values are marked. a: The left-hand bar shows this for all contractions; the right-hand bar shows it only for contractions with a placental volume change exceeding 10%. b: For placental and uterine contractions separately (with a placental volume change > 10% in both cases).

Figure 4a shows that 41% of contractions showed a volume change >10%, and Figure 4b shows that these larger contractions were more likely to be classified as placental (p=0.0061). One contraction was removed from the analysis due to extensive whole maternal movement.

Figure 5 shows contractions defined visually as placental were associated with significantly greater increases in measured placental sphericity and non-placental bed uterine wall area than uterine contractions (p < 0.0001 and p = 0.0015 respectively), with uterine contractions often causing decreases in both these measurements.

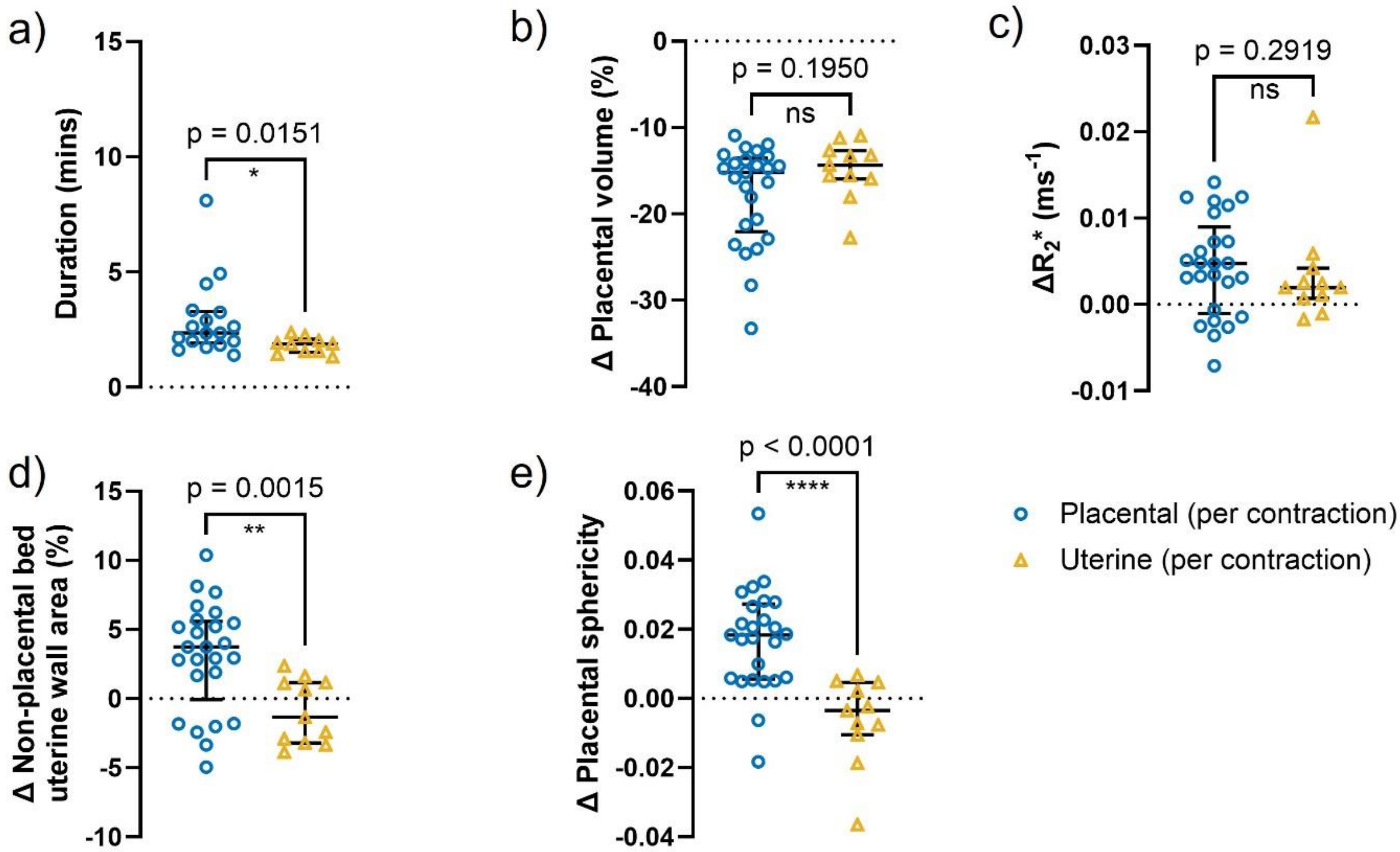

**Fig 5: Characteristics of contractions with volume change >10%, separately for placental and uterine contractions.** a) duration, b) placental volume change, c) change in R2* during a contraction, d) non-placental bed uterine wall area change and e) change in placental sphericity. Median and interquartile range are shown for each group. Results were analysed using the Mann-Whitney U test. Uncorrected p-values are reported, with a significance threshold set at p = 0.01, which was subsequently adjusted to p = 0.002 following Bonferroni correction.

Figure S2 plots these parameters against each other. Figure S2a indicates that there are quantifiable morphological differences between placental and uterine contractions (even for similar changes in net placental volume). Contractions that were likely to be affected by motion are marked in red.

Placental contractions were also longer (p=0.0151) than uterine contractions. There was a non-significant trend for placental contractions to cause a larger increase in $R_2$* and a trend for larger decrease in placental volume.

Figure 6 presents data for all contractions (not just those with >10% placental volume change) and shows that there was large variation both between participants and between contractions from the same individual in a given scan session. It also shows no apparent change with gestational age in the rate and duration of contractions, placental volume reduction or change in $R_2^*$ (p = 0.64, 0.27, 0.65 and 0.56 respectively). There was a trend for the largest %volume changes to occur in early gestation, perhaps reflecting the smaller absolute volume of the resting placenta at that stage of pregnancy. The lack of variation in the change in R2* during a contraction with gestational age is interesting given the known change in baseline R2* of the placenta during pregnancy.

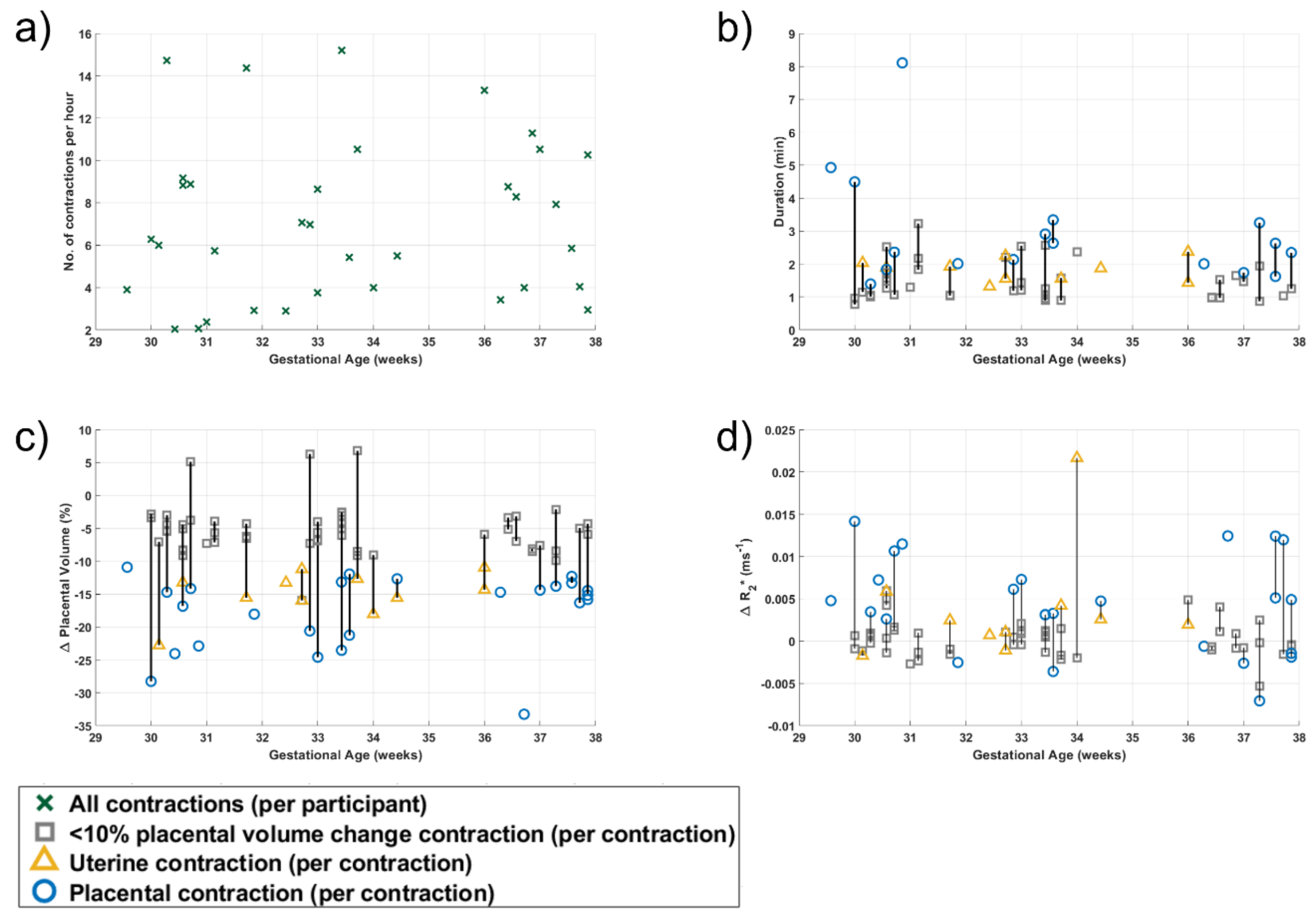

**Fig 6: Variation in contraction features with gestational age for all contractions identified visually (including those with volume changes of less than 10%):** a) rate of any contraction, b) duration, c) change in placental volume and d) change in $R_2^*$ of all contractions across gestational age. Contraction type is shown by marker style (with contractions affected by motion indicated) and contractions from the same individual scan session are joined.

# Discussion

All 35 participants showed placental or uterine contractions over the period for which they were studied, with 30 participants showing any contractions with >10% change in placental volume and 21 showing placental contractions above this threshold. This work supports and characterises the placental contraction phenomenon previously identified *in utero* by our group [8]. Placental contractions observed over a dynamic MRI acquisition of 15-32 minutes, varied in both rate and duration between participants. Placental contractions leading to a volume change >10% occurred at a median rate of two per hour and were observed more often than uterine contractions (Figure 4). This figure does not fully characterise the timing of the contractions, which occur sporadically, sometimes singly and sometimes in a short burst. Longer data acquisition periods, probably with alternative detection mechanisms, are required to investigate the temporal distribution of contractions.

In this work contractions were detected and classified visually by two observers based on morphological features (Figure 3). This subjective approach was necessary given that there is no gold standard at this stage. However, we also identified differences in quantitative metrics between the two types of contractions with the aim of allowing faster and more objective discrimination of contraction types in the future. All contractions were associated with a reduction in placental volume, which is expected since both types will increase the pressure in the uterine cavity. Since the fetus and amniotic fluid are incompressible, both contraction types will push some maternal blood out of the placental intervillous space (IVS) and into the maternal veins, creating this placental volume reduction. There was a trend for placental contractions to involve a larger decrease in volume (up to 35% in some cases), but this difference was not significant, although this may be unreliable as we only considered contractions with >10% placental volume change.

However, there were quantifiable morphological differences between placental and uterine contractions. In particular, the area of the uterine wall not covered by the placental bed increased more in placental than uterine contractions ($p_{adj}$=0.0015, adjusted for 5 comparisons). If placental contractions involve a reduction in the area of the placental bed without any increase in tension in the remaining uterine wall, then they might be expected to lead to the generally observed corresponding increase in the area of the remaining uterine wall, since this must still accommodate the incompressible fetus and amniotic fluid volume. However, in practice, the extent of this increase will depend on whether the reduction in placental volume is sufficient to offset the loss in placental bed area.

Placental sphericity also increased significantly more in placental compared to uterine contractions (p<0.0001). This is likely because when the placenta contracts and the placental bed area gets smaller, the pressure rises in the placenta, and if the rest of the uterine wall can expand somewhat (as discussed above), then the placenta might balloon to support its volume. In contrast, in a uterine contraction the entire uterine wall is under tension, putting the entire contents under pressure, so the placenta cannot balloon in this way. This shape change is discussed further below in the light of knowledge of contractile components in the placenta.

Placental contractions tended to be longer than uterine contractions (p=0.0151, not significant), with some being several minutes in duration. The durations measured using MRI are based on the change in the volume of the placenta which is a determined by both mechanical contraction effects and the haemodynamics of the IVS and vessel refilling, so that these durations cannot be compared directly to shorter durations measured by electrohysterogram (EHG) or mechanical movement of the body wall (detected by the pressure transducer in cardiotocography/CTG). We measured the length as the time between half maximum contraction points, as the full length from the start of contraction to the point of complete refilling is not well defined (Figure 2), and often merged into other events (e.g. other contractions, fetal movement, start or end of data acquisition etc). Nonetheless we did visually estimate the complete lengths to be 3.1 (2.3, 4.7) minutes (median and IQR) and it is likely that the rate of refilling provides information about placental haemodynamics. This is being investigated in future studies including participants with compromised pregnancies.

There was a non-significant trend for placental contractions to be associated with a larger change in placental $R_2$* during the contraction of up to 0.02 $ms^{-1}$ (see also Figure S2). This probably reflects a reduction in spiral artery flow due to the preferential contraction of the placental bed area of the uterine wall, leading to increased deoxygenation during placental contractions. No consistent change in absolute $R_2$* was found between the start and end of a contraction, although since the start and end of the contractions can be difficult to define, this is difficult to assess. Nonetheless it is likely that additional information about the haemodynamic status of the placenta can be obtained from this signal change information.

The objective metrics reported for the different contraction types reflect the features used in visual discrimination and point the way to future automated categorization of contractions, for instance, by thresholding sphericity as seen in the plot shown in Figure S2a. Once the contractions affected by motion are removed, the data suggests that any contraction involving a decrease in volume of more than 10% and increase in sphericity of more than 0.005 could be classified as placental. It is likely that the %change in the non-placental bed area might also inform this classification in future although, as mentioned above, the amount it changes during a placental contraction depends on the relative change in placental volume and placental bed area.

Fetal motion tends to make sphericity more negative, because it generally stretches the uterus making the placenta flatter, and we are investigating whether this can be included in the automatic classification of the contractions. The contractions with small changes in sphericity were those that were more difficult to define visually, which may explain non-zero cut off in sphericity, but we are also investigating whether the visually sorted data can be used to train a machine learning algorithm for detecting and categorising different types of contractions (visual categorisation is not as labour-intensive as a process as manual segmentation).

## Possible mechanisms and implications

Our study confirms the occurrence of the placental contractions that were predicted from prior *work in vitro, in utero,* in healthy pregnancy.

As early as 1906, the term human placenta was noted to have smooth muscle cells in the chorionic plate (Figure 1), and in the 1920s it was the first suggested that the human placenta might be able to contract [15]. During the 1990s Graf et al [16] demonstrated that stem (or anchoring) villi contain myofibroblast cells, and that a second perivascular contractile sheath (PVCS) exists beyond the fetal blood system surrounding each stem villus. They suggested the PVCS could lead to whole placental contraction [17] and hypothesised that it could play a role in both maternal and fetal haemodynamics, including the maintenance of venous return from the placenta.

Subsequently, multiple researchers have reported that contraction of individual stem villi could reproducibly be achieved in the laboratory. Farley et al [18] dissected stem/anchoring villi from term placentae of uncomplicated pregnancies and demonstrated more than 40% increase in contractile tension when they were exposed to potassium chloride, and 75% increase in contractile tension when exposed to a nitric oxide synthetase antagonist l-NAME. They also demonstrated the reversible nature of this phenomenon by exposing contracted villi to sodium nitroprusside and glyceryl trinitrate. These nitric oxide mimetics could induce stem villi relaxation of up to 74%, suggesting a potential role for nitric oxide in the regulation of stem villi contraction/relaxation and therefore materno-fetal blood flow.

Lecarpentier et al [19] used both potassium chloride and electrical tetanus stimulation to assess speed of shortening/contraction and tension in the stem villi. Interestingly, they demonstrated that the speed of stem villi contraction/shortening is the slowest currently known to occur in any mammalian organ, and 20 times slower than smooth muscle contraction in the human uterus. Kato et al[20] built a computational model of stem villi contraction, to demonstrate the likely haemodynamic impact of this phenomenon, showing that stem villi contraction would displace both fetal and maternal circulations. This adds weight to the hypothesis of Farley and Lecarpentier that villous contraction is likely part of the "maternal-fetal blood flow matching" regulatory system present in human placentae.

There have also been previous reports that a myometrial pacemaker exists in the placental bed in rats [21], but the interaction between this and the anchoring villi in humans will require further exploration, possibly including electromyometrial imaging (EMMI) *in utero* [22].

This prior work has largely been laboratory based, so the *in utero* significance of placental contraction mechanisms has remained theoretical, but dynamic MRI offers a method with which to validate this *ex vivo* work and further probe the impact of contractions on maternal and fetal haemodynamics within the placenta.

We observed an increase in sphericity during placental contractions, which is expected if only the placental bed contracts, as the remainder of the uterine wall can expand to accommodate some or all of the contents of the placenta. However, contraction of the stem villi would limit this ballooning and hence ensure that placental contractions squeeze blood from the IVS into the maternal circulation, to cause the observed reduction in placental volume during a placental contraction.

Importantly, we also found that placental contractions were very slow (with a median full duration of three minutes, which is much longer than uterine contractions in labour). This supports Lecarpentier's finding that stem villi show ultra-slow contraction velocity [19].

The underlying physiological significance of these slow placental contractions remains unclear. Contractions clearly refresh the blood in the placenta, which may be expected to increase oxygenation. However, we did not observe a consistent change in $R_2$* between the start and end of the contraction which might suggest there was no bulk change in oxygenation. However, as discussed, this measurement was difficult because the ends of the contractions are hard to define. There was a significant change in $R_2$* during the contraction, but this is to be expected if the inflow and blood volume fraction are reduced and tissue volume fraction is increased at this time. It is notable that the change in $R_2$* during contraction is similar across gestations despite the widely reported drop in baseline placental $R_2$* with gestational age. Given the sensitivity of $R_2$* to oxygenated blood, this suggests a similar haemodynamic response to the contractions across gestation. Further work will consider magnetic susceptibility changes from MRI (which are a more direct measure of oxygenation) and a larger population of women, to allow us to constrain measurements to contractions where the end is clearly defined.

Alternatively, since contractions will disrupt unstirred layers of blood, which are likely to form in the slow flow/low oxygen gradient region of the IVS [8], they may relate to the concept of materno-fetal blood flow matching. There is general consensus that the placenta possesses an intrinsic mechanism to protect fetal oxygen delivery from changes in IVS blood flow, for example during maternal movement [23]. Local flow matching, to maximize efficiency of oxygen delivery, can be achieved either by redistribution of fetal villous blood flow (via vascular cuffs) in response to the oxygen level in the surrounding IVS [24, 25], or by ultra-slow villous/placental bed contractions controlling maternal blood flow into, out of and through the IVS [18, 19]. We are now using mathematical modelling to investigate how to optimize our $R_2$* and more direct oxygenation (magnetic susceptibility) MRI scans to detect such changes.

## Strengths and weaknesses of the study

In order to define how placental contractions contribute to placental physiology, this study investigated placental contractions in a cohort of women with structurally normal and clinically healthy placentas, as confirmed by the ultrasound data obtained during pregnancy, outcomes of labour and birth, and placental histopathology.

However, the study is weakened by its small sample size (35 pregnancies) which will have reduced our statistical power to detect differences in the features of contractions with gestational age or between placental and uterine contractions. In addition, 86% of the women studied were aged over 30 at the time of birth and the average age of our cohort was 32.4 years, whereas the national average age of a first birth is 30.9 years (1). This may reflect the higher proportion (63%) of multiparous women in our study, which was probably because we recruited women with healthy pregnancies, and having had a prior successful pregnancy confers a much lower risk of placental complications in subsequent pregnancies. Another weakness of this study is its low proportion of ethnic minorities. In the UK approximately 18% of the population are non-white and locally in Nottingham this rises to 34% (1). In our cohort only 3 of 35 participants self-reported their ethnicity as non-white, which makes our findings less generalisable to the wider population.

A further limitation of our study is the short duration of the dynamic MRI acquisition (15-32 minutes), limited by the need to ensure the study was acceptable to participants. This snapshot of placental function may not be generalisable over a longer time period, e.g. 24-48 hours, did not allow contractions of both kinds to be seen in all cases, and limited the precision in the measured rate of contractions. Conventional MRI also limited us to studying placental contractions whilst women were lying laterally, which may not be representative of placental function in other positions or during exercise. In future we aim to investigate this using wearable devices.

In our previous work more participants showed no obvious placental contractions during their MRI scan (3), but that was probably because that acquisition was shorter. Furthermore, in the previous study only immediately obvious contractions were manually segmented, whereas in this study full data sets were automatically segmented so that the volume time course curves were available to assist in identifying any contractions. That study was also confounded by the inclusion of hyperoxia in some cases.

Considering technical aspects of this study, the whole volume of the uterus was scanned and segmented to detect placental contractions. This improves the precision of the volume and area time course curves and overcomes the risk of maternal breathing or movement altering the region being interrogated which might create an artificial change in the volume time courses that could be interpreted as a contraction.

In $R_2^*$ weighted MRI, an echo time given by $TE=1/R_2^*$ will provide maximum sensitivity to a change in $R_2^*$, and $R_2^*$ may vary with voxel size as well as field strength. The TE used here was chosen based on data collected previously [8], to provide reasonable sensitivity across both healthy and compromised placentas at different gestational ages. A longer value is likely to be optimal for healthy pregnancies [8, 26] but this would probably provide too little signal at late gestations or in compromised pregnancies. Future work would benefit from optimizing the echo time per subject and possibly working at lower field strength where the $R_2^*$ values will be shorter.

Finally, it would be valuable to acquire simultaneous electrohysterography (EHG/EMG) and MRI data to investigate the electrical signatures of these contractions, although this is challenging since in an MRI scanner physiological abdominal motion induces artefactual signals in EMG data.

## Unanswered questions and future research

Questions remain regarding the physiological trigger initiating placental contractions given the lack of intrinsic pacemaker. In 2010 Suciu et al [27] found telocyte cells in the placental stroma alongside stem villi perivascular contractile sheaths. These telocytes are already known to exist in other endocrine/contractile organs (the small intestine and pancreas). They proposed that these telocytes represent a co-ordinated network across the placenta which allows stem villi to contract in a co-ordinated fashion. Electrophysiological currents generated by telocytes [28] are capable of initiating smooth muscle contraction, for example in the myometrium of the uterus (raising the possibility that the placenta could provide a uterine pacemaker in pregnancy), but as the placenta is a non-innervated organ, other researchers have suggested that co-ordination of villi contraction may be by direct cell to cell communication or via a paracrine mechanism [29]. However, the pacemaker could be in the placental bed and then spread to the villi. It is possible that carefully designed in vivo experiments, including looking at the time courses of the different changes, may be able to answer this question.

In this work, we have demonstrated that placental contractions are accompanied by a reduction in the placental bed area of the uterine wall and the volume of the placenta, but we cannot directly observe the behaviour of the villi. However, if the stem villi and placental bed did not contract in concert, then a reduction in placental bed area alone would be inefficient and would lead to excessive ballooning of the chorionic plate depending on the tension in the rest of the uterine wall. Direct imaging of the villous trees, and a mechanical and hydrodynamic model of the placenta using MRI time courses of placental structural changes as inputs, might provide ways to test whether contractions are triggered by the placental bed or villi.

We now need to investigate whether placental contractions are altered in compromised pregnancies. In 2010 Khozhai et al[30] examined the difference in number of smooth muscle cells/myofibroblasts between healthy and compromised placentas, and found more extravascular stromal smooth muscle cells and thicker perivascular contractile sheaths in pre-eclamptic and growth-restricted pregnancies. What remains unclear is whether the increase in extravascular contractile cell types is the result of placental dysfunction or whether it generates haemodynamic changes which lead to placental dysfunction. Suciu et al [27] hypothesised a link with placental pathology/insufficiency/remodelling as telocytes are known to express angiogenic vascular endothelial growth factor (VEGF), a biomarker whose role in pre-eclampsia is yet to be fully established. Bosco et al [29] reviewed the role of telocytes in co-ordinating stem villi contraction and therefore materno-fetal blood flow and hypothesized that the oxidative stress resulting from maternal under-perfusion could reduce telocyte and therefore pacemaker function, which in turn could disrupt stem villi contraction and negatively impact oxygen and nutrient transfer across the IVS. However, further work is required to understand the role of the contractions in healthy pregnancies and those compromised by conditions such as fetal growth restriction or pre-eclampsia.

These findings have informed the design of an ongoing longitudinal study into the characteristics of placental contractions in compromised pregnancies, through which we aim to establish if there are associations between the rate, duration or nature of placental contractions with other markers of placental or fetal wellbeing or with adverse pregnancy outcomes. Furthermore, we are developing a wearable device which will allow us to monitor placental contractions in varying maternal positions and over a longer time period to improve our understanding of this phenomenon.

# Conclusion

This study extends our previous work, providing further *in utero* evidence for placental contractions, supporting previously predictions from *in vitro* studies. We have also provided quantitative metrics (based on automatic segmentation of the MRI scans) which can be used to discriminate placental from uterine contractions in healthy pregnancy, which will accelerate future *in utero* work exploring this area.

# Supporting Information

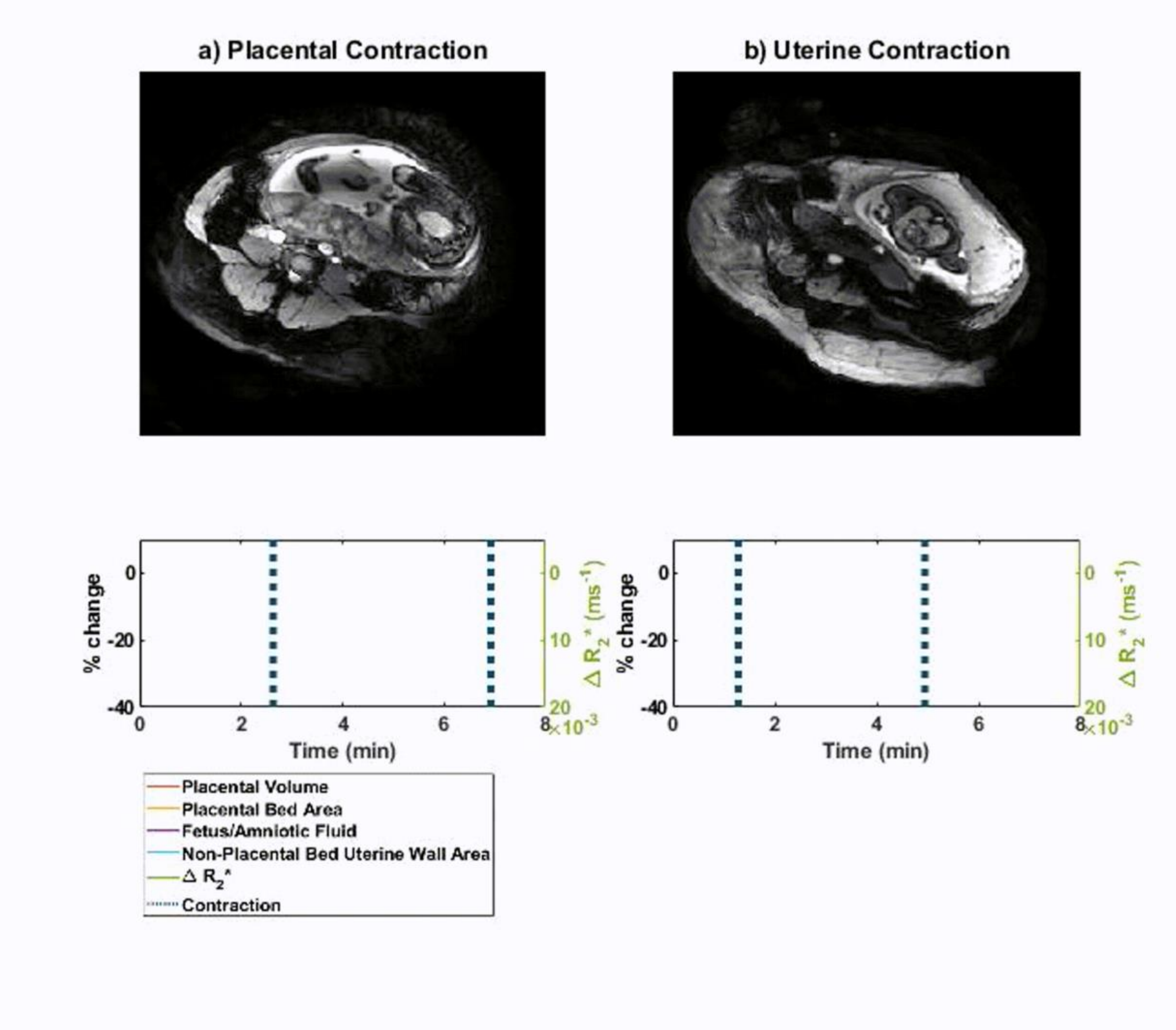

**S1: Dynamic MRI video of contractions: a) placental and b) uterine contraction with associated changes in volume/area measurements and $\Delta R_2$* of the placenta.**

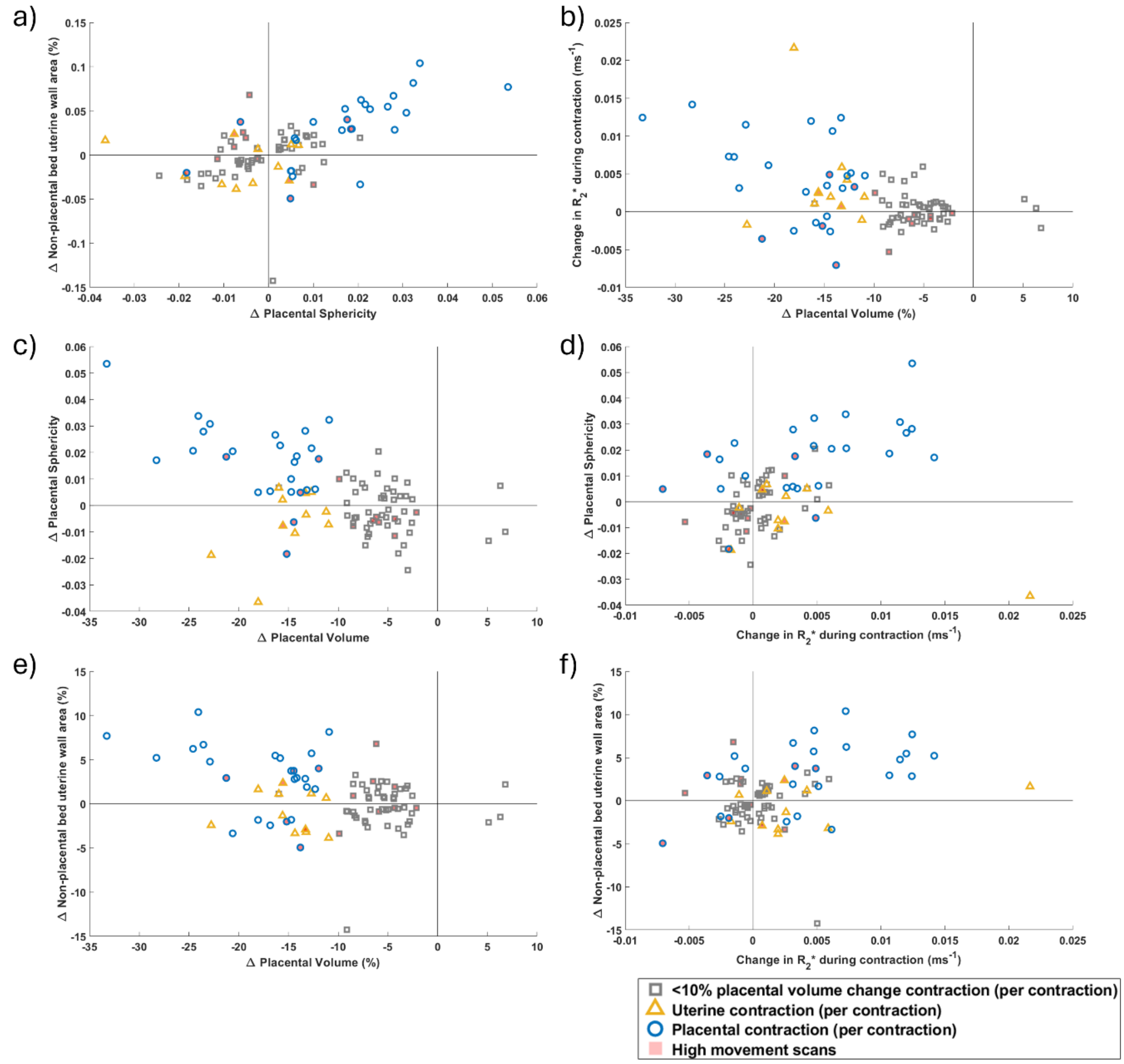

**Fig S2: Relationship between change in placental sphericity and change in non-placental bed uterine wall area against contraction features per contraction, indicating morphological differences between contraction types.** a) change in placental sphericity against change in non-placental bed uterine wall area, change in placental volume against b) change in placental sphericity and c) non-placental bed uterine wall area, change in $R_2$* against d) change in placental sphericity and e) non-placental bed uterine wall area.